\documentclass[groupedaddress,showkeys]{revtex4-2}
\usepackage{float}
\usepackage{multirow}
\usepackage{amsmath}
\usepackage{amssymb}
\usepackage{graphicx}
\usepackage{stackrel}
\usepackage{xcolor}
\usepackage{natbib}
\definecolor{tl}{RGB}{0,180,120}
\usepackage[colorlinks,citecolor=tl,linkcolor=red,urlcolor=magenta,unicode=true]{hyperref}

\begin{document}
	\title{Exploring the effects of dark matter - dark energy interaction on cosmic evolution in viscous dark energy scenario}
	
	\author{Ashadul Halder}
	\email{ashadul.halder@gmail.com}
	\affiliation{ Department of Physics, St. Xavier's College,\\ 30, Mother Teresa Sarani, Kolkata 700016, India. }
	
	\author{Madhurima Pandey}
	\email{madhurima0810@gmail.com}
	\affiliation{Department of Physics, School of Applied Sciences and Humanities, \\Haldia Institute of Technology, Haldia, 721657, West Bengal, India.}
	
	\author{Rupa Basu}
	\email{rupabasu.in@gmail.com}
	\affiliation{Department of Physics, St. Xavier's College,\\ 30, Mother Teresa Sarani, Kolkata 700016, India. }
	
	\author{Debasish Majumdar}
	\email{debasish.majumdar@gm.rkmvu.ac.in}
	\affiliation{Department of Physics Ramakrishna Mission Vivekananda Educational and Research Institute, \\Belur Math, Howrah 711202, West-Bengal, India}

	\date{\today}
	\begin{abstract}
		We explore the influence of interactions between dark matter \textbf{(DM)} and dark energy \textbf{(DE)} on the cosmic evolution of the Universe within a viscous dark energy (VDE) framework. Moving beyond traditional interacting dark energy (IDE) models, we propose a generalized IDE model adaptable to diverse IDE scenarios via IDE coupling parameters. In order to investigate deviations from $\Lambda$CDM across cosmic epochs by highlighting how viscous and the interactions between DM and DE impact cosmic density and expansion rates, we consider a model agnostic form of VDE.  Eventually we perform a Bayesian analysis using the Union 2.1 Supernova Ia dataset and Markov Chain Monte Carlo (MCMC) sampling to obtain optimal values of model parameters. This comprehensive analysis provides insights about the interplay between viscous and IDE in shaping the Universe's expansion history.
		
	\end{abstract}
	\maketitle
	

	\section{Introduction}
	There is no denying that a new data driven age in modern cosmology has begun. In 1998, Supernova Type Ia (SNIa) \cite{Riess_1998,PhysRevLett.83.670} observation first discovered that the Universe is not only expanding, but also accelerating. Subsequently, the phenomenon of the Universe's late time acceleration have been confirmed by several studies such as the measurement of baryon acoustic oscillations (BAO) \cite{2013PhR...530...87W} in the galaxy power spectrum, large scale structure observation and Cosmic Microwave Background Radiation (CMBR) \cite{Bennett_2013,2014A&A...571A..16P} observations. The known luminous object in the Universe accounts for only about 4.9\% of the total mass energy budget of the Universe. The analysis of satellite borne PLANCK experiment on the anisotropy of CMBR suggests that the Universe overwhelmingly contains unknown dark matter (DM) and dark energy (DE). The former one is an unseen unknown matter, whose only definite evidence so far is due to their gravitational effects, makes up for around 26.8\% of the Universe’s contents. The rest around 69.3\% is yet another unknown energy known as DE thought to invoke a negative pressure opposite to gravity and cause the recent accelerated expansion of the Universe.
	
	As previously mentioned, the pressure acts in opposition to gravity to accelerate the Universe against the propensity of an eventual gravitational collapse caused by the mass contained within the Universe. Since then, a number of theoretical frameworks have been proposed to explain the nature and origin of the mysterious DE and the ensuing late temporal acceleration. The cosmological constant scenario, which is popularly known as $\Lambda$CDM model, is considered to be a DE candidate. Recent PLANCK CMB measurements have validated the $\Lambda$CDM model as a phenomenologically feasible cosmological model \cite{2016A&A...594A..13P}. However, this well known model ($\Lambda$CDM model) is insufficient to fully describe our Universe \cite{RevModPhys.61.1}. The physical properties of DE and DM are still an enigma from a theoretical perspective. The cosmological constant has long standing coincidence and fine tuning issues \cite{RevModPhys.61.1,PhysRevLett.82.896} . According to observations, the $\Lambda$CDM model can satisfactorily explain 
	cosmic large - scale structures, but it encounters several challenges at the dwarf - galaxy scale, including the well known issues with missing satellites, core - cusp and too - big - to - fail problems \cite{2017ARA&A..55..343B}. A few disparities between various observations under $\Lambda$CDM have appeared and become increasingly significant in the last few years. The most prominent one is the $H_0$ tension, obtained from Planck CMB data, which is around 5$\sigma$ apart from the values acquired from standard candles and lensing time - delay observations \cite{Riess:2019qba,Riess_2019,2020MNRAS.498.1420W}. The other less important one is the $S_8$ tension having a 1 - 2$\sigma$ differ between the value of $S_8 = \sigma_8 (\Omega_m)^{0.5}$ bounded by Planck CMB and that from weak lensing measurements \cite{2019PASJ...71...43H,2020A&A...633A..69H,2021A&A...645A.104A,PhysRevD.105.023514}. 
	
	Since, we still don't completely understood how realistic the DE is, cosmologists have put out a plethora of options to address these mysteries. In general, we can classify them into two major categories, i.e., the DE models \cite{CALDWELL200223,KAMENSHCHIK2001265} and modified theory of gravity. Within the context of general relativity (GR) \cite{PhysRevD.68.123512,Dvali:2000hr}, the former one seeks to introduce a new cosmic fluid or matter field. However, in the latter scenario, the standard GR lagrangian is modified by a physical mechanism that is specific to the breakdown of Einstein's gravity. 
	
	From thermodynamics it can be stated that, a Universal feature of any realistic fluid dynamics is dissipative process. This dissipative process has been taken into account to conduct several cosmological analyses. Both bulk and shear viscosity in the stress energy tensor of the cosmic fluid \cite{PhysRev.58.919,PhysRevD.63.023501} are furnished by such dissipative process. Although the shear viscosity can be disregarded at large scales in the case of isotropic and homogeneous Universe, but the impact of bulk viscosity has been widely studied in the context of viscous dark matter (VDM) \cite{PhysRevD.64.063501,PhysRevD.86.083501}, viscous dark energy (VDE)  \cite{REN20061,MengXinHe2009,PhysRevD.76.103516,2016arXiv161101023K} and cosmic inflation. Additionally, in various models, viscosity has been suggested as a possible explanation for the current acceleration. Viscosity not only modifies  the Hubble evolution, but it also produces additional entropy that warms the baryon and dark matter fluids that may affect the thermal evolution of the Universe.

	In the present work, the possible impact of the DM and DE interaction in VDE framework on the dynamics of the Universe has been addressed. In lieu of considering different interacting dark energy (IDE) models \cite{idem0,idem1,21cm_ide}, we introduce an generalized IDE model in our calculation. Different IDE models can be obtained from this generalized model by tuning its parameters. The evolution equation of Hubble parameters due to different VDE models can also be acquired from a generalized expression as introduced in \cite{21cm_vde}. In our analysis, we investigate the dynamics of the Universe is studied by studying evolution of cosmological density parameters and higher order evolution parameters namely, deceleration parameter ($q$), jerk parameter ($j$) and snap parameter ($s$). In order to understand the variation of the Universe's mass fraction over time, an extensive study has been done on $Om$ parameter in the present work. We further extend our analysis using three point diagnostic $Om3$ parameter as described in \cite{sahni}.
	
	The paper is organized as follows. In \autoref{sec:DMDE} we discussed about interplay between DM and DE. Different VDE models and their effect on the evolution of Hubble parameter have been discussed extensively in \autoref{sec:VDE}. Calculations and results are shown in \autoref{sec:calc}. Eventually, in \autoref{sec:conc} some concluding remarks are given.
	
	\section{Interplay Between Dark Matter and Dark Energy} \label{sec:DMDE} 
	The interaction between DM and DE could significantly influence universal dynamics, including the optical depth and spin temperature of the 21-cm transition. In the standard cosmological model, the density parameters of DM ($\Omega_{\chi}$) and DE ($\Omega_{\rm de}$) evolve as $\Omega_{\chi,0}(1+z)^3$ and $\Omega_{\rm de,0}(1+z)^{3(1+\omega_{\rm de})}$ respectively, where $\Omega_{\chi,0}$ and $\Omega_{\rm de,0}$ are the respective density parameters at the redshift $z=0$, and $\omega_{\rm de}$ is the equation of state (EOS) parameter of DE. However, the DM and DE densities, in the presence of their in between interactions, have been evolved in the following manner \cite{Li}:
		 	
	\begin{equation}
		(1+z) H(z) \frac{{\rm d} \rho_{\chi}}{{\rm d} z}-3H(z)\rho_{\chi}  = -\mathcal{Q} 
		\label{eq:rho_chi}
	\end{equation}
	\begin{equation}
		(1+z) H(z) \frac{{\rm d} \rho_{\rm de}}{{\rm d} z} - 3 H(z) (1+ \omega_{\rm de}) \rho_{\rm de} = \mathcal{Q}
		\label{eq:rho_de}
	\end{equation}
	
	Here, $\mathcal{Q}$ denotes the energy transfer between DM and DE due to DM-DE interaction. In this work, we consider three benchmark models to investigate the effect of DM-DE interaction on brightness temperature. The energy transfer expressions for those benchmark models are described below \cite{model1,model2,model3,model4}:
	
	\begin{center}
		\begin{tabular}{ll}
			Model-I \hspace{5mm} & $\mathcal{Q}=3 \lambda H(z) \rho_{\rm de}$\\
			Model-II & $\mathcal{Q}=3 \lambda H(z) \rho_{\chi}$\\
			Model-III &$\mathcal{Q}=3 \lambda H(z) (\rho_{\rm de} +\rho_{\chi})$
		\end{tabular}
	\end{center}
	
	Here, $\lambda$ is the coupling parameter, determining the strength of the DM - DE interaction and $\rho_{\rm de}$, $\rho_{\chi}$ are the densities of DE and DM respectively. The stability conditions for each of the models are described in Table~\ref{tab:stability}. Various phenomenological studies have constrained these models using observational data from PLANCK, SNIa, and BAO \cite{model3, model4, model_benchmark1, model_benchmark2, model_benchmark3,
		model_benchmark4, model_benchmark5, model_benchmark6}.
	
	\begin{table}
		\centering
		\caption{\label{tab:stability} Stability conditions of the model parameters for different IDE models}
		\begin{tabular}{lccr}
			\hline
			Model & $\mathcal{Q}$ & EOS of dark energy & Constraints\\
			\hline
			I & $3 \lambda H(z) \rho_{\rm de} $ & $\omega_{\rm de}<-1$ & $\lambda<- 2 \omega_{\rm de} \Omega_{\chi}$\\
			II & $3 \lambda H(z) \rho_{\chi} $ & $\omega_{\rm de}<-1$ & $0<\lambda<-\omega_{\rm de}/4$\\ 
			III & $3 \lambda H(z) (\rho_{\rm de} + \rho_{\chi}) $ & $\omega_{\rm de}<-1$ & $0<\lambda<-\omega_{\rm de}/4$\\
			\hline
		\end{tabular}
		
	\end{table}
	
	All these three models can be expressed using a general expression given by,
	\begin{equation}
		\mathcal{Q}=3 H(z) (\lambda_{\rm de}\rho_{\rm de} +\lambda_{\chi} \rho_{\chi}).
	\end{equation}
	
	In the above equation, $\lambda_{\rm de} (\lambda_\chi)$ indicated the coupling parameter for DE (DM). 
	In the present analysis we parameterize the coupling parameters $\lambda_{\chi}$ and $\lambda_{\rm de}$ in terms of $\lambda_r$ and $\lambda_{\theta}$, as 
	\begin{eqnarray}
		\lambda_{\chi}   & = & \lambda_r\,\sin(\lambda_{\theta})\\
		\lambda_{\rm de} & = & \lambda_r\,\cos(\lambda_{\theta})
	\end{eqnarray}
	
	\section{Viscous Dark Energy (VDE) Models} \label{sec:VDE}  
	The dynamics of the Universe may be significantly impacted by the presence of viscosity in the dark energy fluid. The two Freidmann equations that govern the dynamics of Friedmann - Robertson - Walker (FRW) Universe are as follows
	\begin{eqnarray}
		\dfrac{\dot{a}^2}{a^2} &=& \dfrac{\rho}{3}, \\
		\dfrac{\ddot{a}}{a} &=& -\dfrac{\rho +3p}{6}. \label{eq:1}
	\end{eqnarray}
	In the above equations, the dot represents the derivative with respect to the cosmic time $t$, whereas the parameters $a$, $\rho$ and $p$ stand for the scale factor, energy densities and pressure of the cosmic fluids, which include relativistic radiation, baryon matter, DM and DE, respectively. Throughout this work, we consider $8\pi G = c = \hbar = 1$. The stress - energy tensor of a cosmic DE fluid with the bulk viscosity $\zeta$ is expressed as
	\begin{equation}
		T_{\mu \nu}^{\rm de} = \rho_{\rm de} U_\mu U_\nu + \overline{p_{\rm de}} h_{\mu \nu}, \label{eq:2}
	\end{equation} 
		where $\overline{p_{\rm de}}$ and $\rho_{\rm de}$ indicate the effective pressure and the energy density of the VDE fluid respectively. $U_\mu$ is the four - velocity of the VDE fluid in comoving coordinates, where $U_\mu = (1,0,0,0)$ and the projection tensor is denoted by $h_{\mu \nu} = g_{\mu \nu} + U_\mu U_\nu$. According to Eckart's theory as a first order limit of the Israel - Stewart scenario having zero relation time, the effective pressure $\overline{p_{\rm de}}$ at thermodynamical equilibrium for a typical VDE fluid can be restated as \cite{PhysRev.58.919}
		\begin{equation}
			\overline{p_{\rm de}} = \omega_{\rm de}\rho_{\rm de} = \zeta \theta. \label{eq:3}
		\end{equation}
	In the above equation, $\omega_{\rm de}$, $\zeta$ and $\theta$ denote the equation of state (EoS) of DE fluid, bulk viscosity and expansion scalar respectively. Moreover, the energy conservation equation associated with the VDE fluid can be expressed as 
	\begin{equation}
		\dot{\rho_{\rm de}} + \theta (\overline{p_{\rm de}} + \rho_{\rm de}) = 0, \label{eq:4}
	\end{equation}
	where $\theta = 3H = \dfrac{\dot{a}}{a}$ and $H$ defines the Hubble parameter, which governs the background evolution of the Universe. 
	
	In our analysis, we assume that the bulk viscosity of DE is proportional to the Hubble parameter as $\zeta = \eta H$, where $\eta$ is a dimensionless free parameter. Therefore, the effective pressure $\overline{p_{\rm de}}$ in \autoref{eq:3} turns out as 
	\begin{eqnarray}
		\overline{p_{\rm de}} &=& \omega_{\rm de}\rho_{\rm de} - 3\eta H^2. \label{eq:5}
	\end{eqnarray}
	
	The effects of three VDE models have been investigated in our analysis. 
	
		$\bullet$ {\bf Model I (V$\Lambda$DE)}: A one parameter extension of the conventional six parameter $\Lambda$CDM model has been taken into account in this first model. Here $\omega_{\rm de}$ is fixed at -1. Therefore, we refer this model as the V$\Lambda$DE model and its effective pressure takes the form $\overline{p_{\rm de}} = - \rho_{\rm de} - 3 \eta H^2$. The dimensionless Hubble parameter $H(z)$ of the V$\Lambda$DE model is given by
		\begin{equation}
		H(z) = H_0 \left[\dfrac{1}{1+\eta}\Omega_{m,0} (1+z)^3 + 
		\left(1-\dfrac{1}{1+\eta}\Omega_{m,0} \right) (1+z)^{-3\eta}\right]^{1/2},\label{eq:mod1} 
	\end{equation}
	where the present value of the Hubble parameter is denoted as $H_0$. In the above equation (\autoref{eq:mod1}), $z$ and $\Omega_{m,0}$ represent the redshift and current density parameter of matter including both baryonic matter and DM. It is to be noted that, the contribution from the radiation and spatial curvature components have been ignored in this model. If we put $\eta = 0$ in \autoref{eq:mod1}, then V$\Lambda$DE model will be reduced to the standard $\Lambda$CDM model.
	
	$\bullet$ {\bf Model II (V$\omega$DE)}: In this model, an additional free parameter $\omega_{\rm de}$ is introduced along with the viscosity parameter $\eta$. Therefore, this model has been mentioned as V$\omega$DE model. Furthermore, the Hubble parameter $H(z)$ in V$\omega$DE model can be expressed as 
	\begin{equation}
	H(z) = H_0 \left[\dfrac{\omega_{\rm de}}{\omega_{\rm de}-\eta}\Omega_{m,0} (1+z)^3 + \left(1-\dfrac{\omega_{\rm de}}{\omega_{\rm de}-\eta}\Omega_{m,0} \right) (1+z)^{3(1+\omega_{\rm de}-\eta)}\right]^{1/2}.\label{eq:mod2}	
	\end{equation}
	$\bullet$ {\bf Model III (VKDE)}: In order to account for the impact of bulk viscosity on the spatial curvature, we incorporate the curvature contribution into the cosmic pie. The present vakue of the curvature density parameter $\Omega_k$ is considered as another model parameter. We can exhibit the corresponding Hubble parameter of the VKDE model as 
	\begin{eqnarray}
		H(z) &=& H_0 \left[\dfrac{2}{2+3\eta}\Omega_{k,0} (1+z)^2 + 
		\dfrac{1}{1+\eta}\Omega_{m,0} (1+z)^3 + \right.\nonumber\\ &&\left.\left(1-\dfrac{2}{2+3\eta}\Omega_{k,0}-\dfrac{1}{1+\eta}\Omega_{m,0} \right) (1+z)^{-3\eta}\right]^{1/2}\label{eq:mod3},
	\end{eqnarray}
	where $\Omega_{k,0}$ is the current curvature density parameter. In \autoref{tab_mod}, we tabulate the constraints and parameters of the three individual models. 
	Since bulk viscosity has minimal effect at low redshifted epochs, the evolution of Universe closely resembles the standard $\Lambda$CDM cosmology for lower redshifts. The bulk viscosity of DE is addressed by th three previously mentioned VDE models. At lower redshift, the second model offers a relatively faster expansion of the Universe for $\omega_{\rm de} \leq -1$, as an extra effect of $\omega_{\rm de}$ (EoS of DE) is considered. The expansion rate, however, sharply declines for $\omega_{\rm de} \gtrapprox -0.7$. In contrast the third VDE model investigates the impact of the Universe's spatial curvature in the presence of viscous flow of DE. It can be observed that the evolution of the Universe is minimally modified. at late times according to the recent bounds of curvature. 
	
	\begin{table}
		\centering
		\begin{tabular}{ccc}
			\hline
			Model & Constraints & Free parameters\\
			\hline
			I & $\quad$ $\omega_{\rm de}=-1$, $\Omega_k=0$ $\quad$ & $\eta$\\
			II & $\Omega_k=0$ & $\eta$, $\omega_{\rm de}$\\
			III & $\omega_{\rm de}=-1$ & $\eta$, $\Omega_k$\\
			\hline
		\end{tabular}
		\caption{\label{tab_mod} Constraints and parameters of three viscous dark energy models}
	\end{table}
	
	The generalized expression of the Hubble parameter for all three models (\autoref{eq:mod1}, \autoref{eq:mod2}, \autoref{eq:mod3}) can be written as
	\begin{eqnarray}
		H(z)&=&H_0 \left[\dfrac{2}{2+3\eta}\Omega_{k,0} (1+z)^2 +
		\dfrac{\omega_{\rm de}}{\omega_{\rm de}-\eta}\Omega_{m,0} (1+z)^3\right.\nonumber \\
		&&\left. + \left(1-\dfrac{2}{2+3\eta}\Omega_{k,0}-\dfrac{\omega_{\rm de}}{\omega_{\rm de}-\eta}\Omega_{m,0} \right) (1+z)^{3(1+\omega_{\rm de}-\eta)}\right]^{1/2}.
		\label{eq:general_H}
	\end{eqnarray}
	In the case of interacting dark energy model, the above expression takes the form,
	\begin{equation}
		H(z)=H_0 \left[\dfrac{2}{2+3\eta}\Omega_{k,0}(1+z)^2+\dfrac{\omega_{\rm de}}{\omega_{\rm de}-\eta}\left(\Omega_{b,0}(1+z)^3+\dfrac{\rho_{\chi}}{\rho_{c,0}}\right)+(1+z)^{-3\eta}\dfrac{\rho_{\rm de}}{\rho_{c,0}}\right]^{1/2}.
	\end{equation}

	\section{Calculations and Results} \label{sec:calc}
	
	How the interactions between DM and DE impact the evolution of the Universe within a VDE framework has been explored in this work. Rather than relying on conventional IDE models, we introduce a generalized IDE model that can represent a wide array of IDE scenarios by tuning its parameters, $\lambda_r$ and $\lambda_{\theta}$. By adopting a generalized form of the Hubble parameter in VDE framework, we establish a basis for our calculations. This form enables us to systematically analyze the dynamic effects of viscosity and dark sector interactions on cosmic evolution, facilitating a versatile approach to studying different IDE conditions within a unified framework.
	
\begin{figure}
		\centering
		\includegraphics[width=0.48\textwidth]{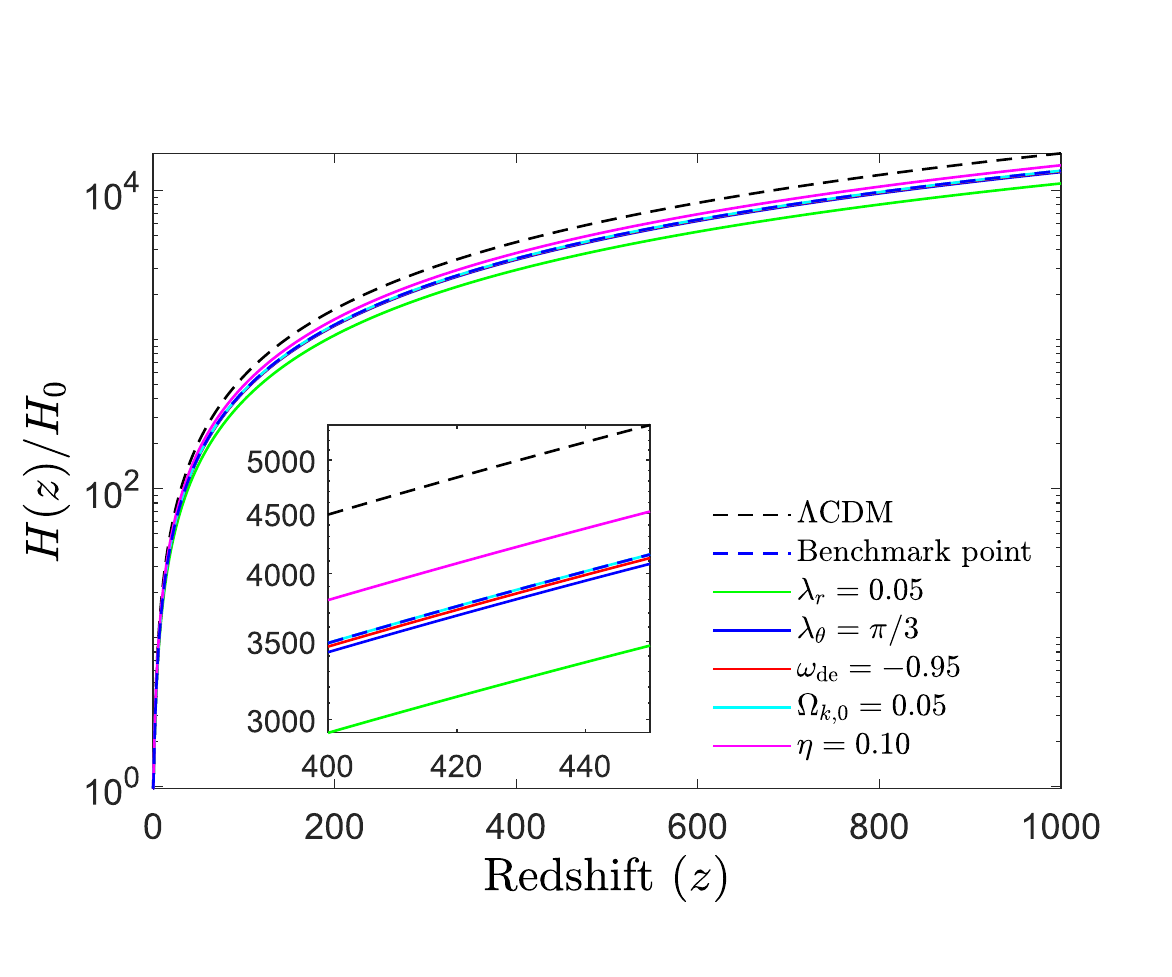}
		\caption{\label{fig:first}Variation of Hubble evolution ($H(z)/H_0$) with redshift $z$ for different model parameters. See text for detail.}
	\end{figure}

	In \autoref{fig:first}, the Hubble evolutions with redshift $z$ are shown for different chosen parameter values. In this regard, we consider a set of benchmark values of the model parameters as shown in \autoref{tab_bp}. The Hubble evolution corresponding to these benchmark values (blue dashed line) exhibits slightly lower values of $H(z)/H_0$ for different redshifts $z$ compared to that of the $\Lambda$CDM model (black dashed line). It is to be noted that for the $\Lambda$CDM model, we choose $h_0=0.674$. The green solid line represents the same for $\lambda_r=0.05$, while keeping the values of the other parameters the same as in \autoref{tab_bp}. Similarly, the blue solid line, red solid line, cyan solid line, and magenta solid line describe the Hubble evolution with redshift $z$ for $\lambda_{\theta}=\pi/3$, $\omega_{\rm de}=-0.95$, $\Omega_{k,0}=0.05$, and $\eta=0.1$, respectively, while keeping the other parameters fixed at benchmark values for individual cases. From this figure, it can be seen that the effect of $\Omega_{k,0}$ is minimal on the evolution of the Hubble parameter.
	
	\begin{table}
		\centering
		\begin{tabular}{c c c c c c}
			\hline\hline
			$\quad h_0 \quad$ & $\quad\lambda_r \quad$ & $\quad \lambda_{\theta} \quad$ & $\quad \omega_{\rm de} \quad$ & $\quad \Omega_{k,0} \quad$ & $\quad \eta \quad$\\ \hline
			0.674 & 0.02 & $\pi/4$ & -1.0 & 0.01** & 0.3\\
			\hline\hline
		\end{tabular}
		\caption{\label{tab_bp} Benchmark point for different parameter for \autoref{fig:first}, \ref{fig:Omega} and \ref{fig:hps}.}
	\end{table}
	
\begin{figure*}
		\centering
		\begin{tabular}{cc}
			\includegraphics[width=0.49\textwidth]{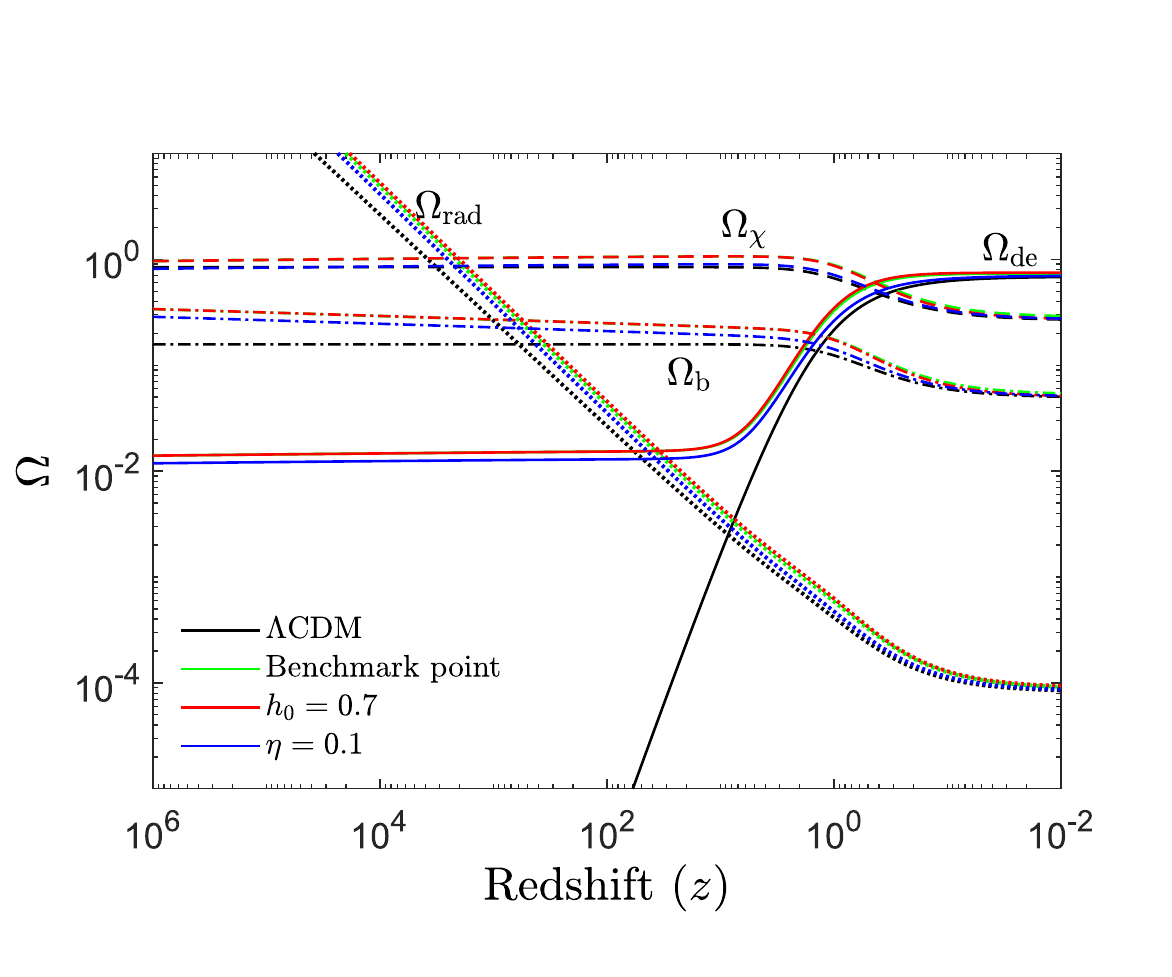}&
			\includegraphics[width=0.49\textwidth]{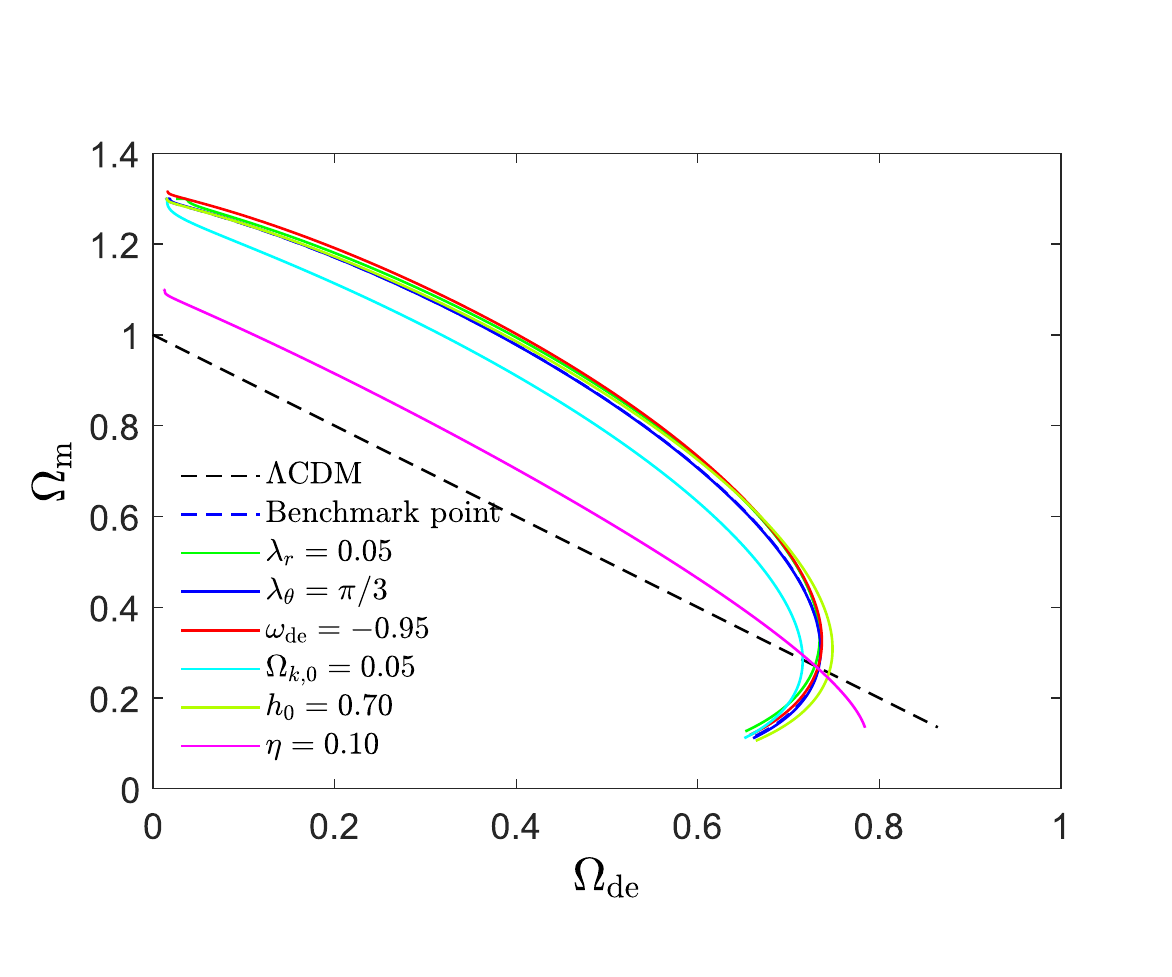}\\
			(a)&(b)\\
		\end{tabular}
		\caption{\label{fig:Omega}(a) Evolution of density parameters with redshift $z$ for different model parameters. See text for detail. (b) Parametric representation of $\Omega_{\rm de}$ and $\Omega_{\rm m}$ ($\Omega_{\chi}+\Omega_{\rm b}$). See text for detail.}
	\end{figure*}

	In presence of VDE and interaction between DM and DE, the evolution history of the density parameters are remarkably different from the $\Lambda$CDM model (see \autoref{fig:Omega}(a)). In \autoref{fig:Omega}(a), the black and green solid lines represent the evolution of dark matter density parameter $\Omega_{\rm de}$ for the case of $\Lambda$CDM model and the present model with the benchmark values of model parameters (see \autoref{tab_bp}). The red and blue solid lines represent the same for the benchmark values with additional modifications $h_0=0.7$ and $\eta=0.1$ respectively. The dashed, dash-dotted and dotted lines of corresponding colour exhibits the evolution of dark matter density parameter ($\Omega_{\chi}$), baryon density parameter ($\Omega_{\rm b}$) and radiation density parameter ($\Omega_{\rm rad}$) respectively for individual cases. From this figure one can see that, at the present epoch the values of density parameters are independent to model and model parameters, however the dependence on model and model parameters become prominent as we approach toward higher values of $z$. Here, the effect of $h_0$ is prominent in the case of $\Omega_{\rm rad}$ only, while it is negligible in the cases of dark matter,dark energy and baryons. On the other hand, the density parameters for all constituents of the Universe show notable variation for $\eta=0.1$.
	
	We also show a parametric representation of dark energy density parameter $\Omega_{\rm de}$ and matter density parameter $\Omega_{\rm m}$, i.e. $\Omega_{\chi}+\Omega_{\rm b}$ in \autoref{fig:Omega}(b). Here the $\Lambda$CDM model demonstrate a straight line (black dashed line), while in presence of the interactions in dark sector components, the representative lines become curved. From the magenta line of this figure (\autoref{fig:Omega}(b)) one can conclude that the bulk viscosity parameter $\eta$ plays a crucial role in the estimation of the curvature of that line.
	
	To achieve a thorough comprehension of the cosmological evolution, it is essential to ascertain the following parameters \cite{qjs_param}:
	\begin{eqnarray}
		q &=& -\dfrac{1}{a}\dfrac{{\rm d}^2 a}{{\rm d} t^2} \left[\dfrac{1}{a} \dfrac{{\rm d} a}{{\rm d} t}\right]^{-2},\\
		j &=& +\dfrac{1}{a}\dfrac{{\rm d}^3 a}{{\rm d} t^3} \left[\dfrac{1}{a} \dfrac{{\rm d} a}{{\rm d} t}\right]^{-3},\\
		s &=& +\dfrac{1}{a}\dfrac{{\rm d}^4 a}{{\rm d} t^4} \left[\dfrac{1}{a} \dfrac{{\rm d} a}{{\rm d} t}\right]^{-4},\\.		
	\end{eqnarray}
	In the above expressions, $q$ denotes the deceleration parameter of the Universe. It can alternatively be expressed as $q = -\frac{\ddot{a}a}{\dot{a}^2}$, where $a$ denotes the scale factor, and the over-dot signifies the time derivative. Additionally, the parameter $j$ represents the jerk parameter, which characterizes the rate of change of the deceleration parameter with time. The snap parameter $s$ represents the rate of change of the jerk parameter with time and the $Om$ parameter provides a measure of the cosmic acceleration relative to the gravitational attraction exerted by matter. 
	
	\begin{figure*}
		\centering
		\begin{tabular}{cc}
			\includegraphics[width=0.49\textwidth]{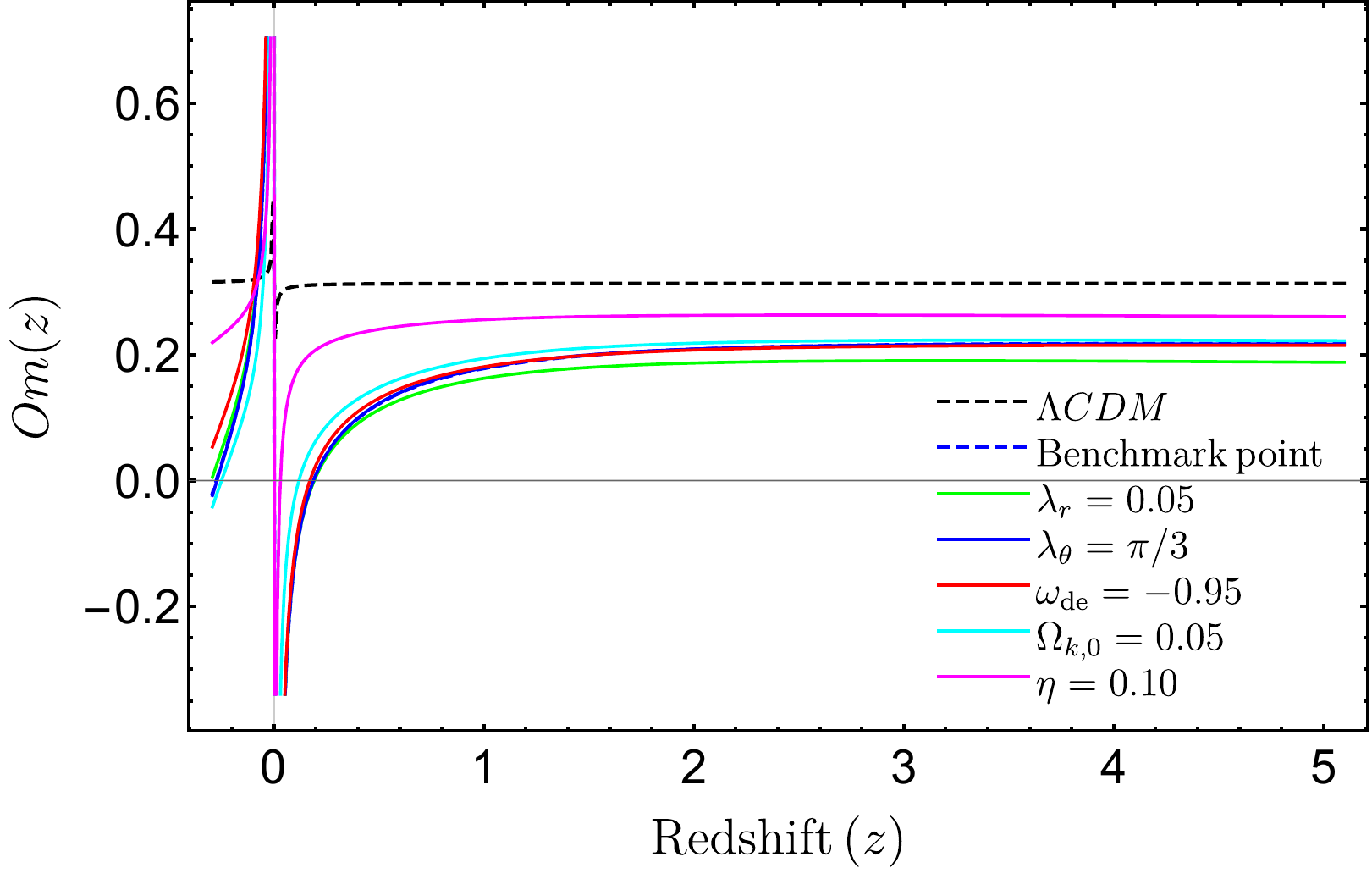}&
			\includegraphics[width=0.49\textwidth]{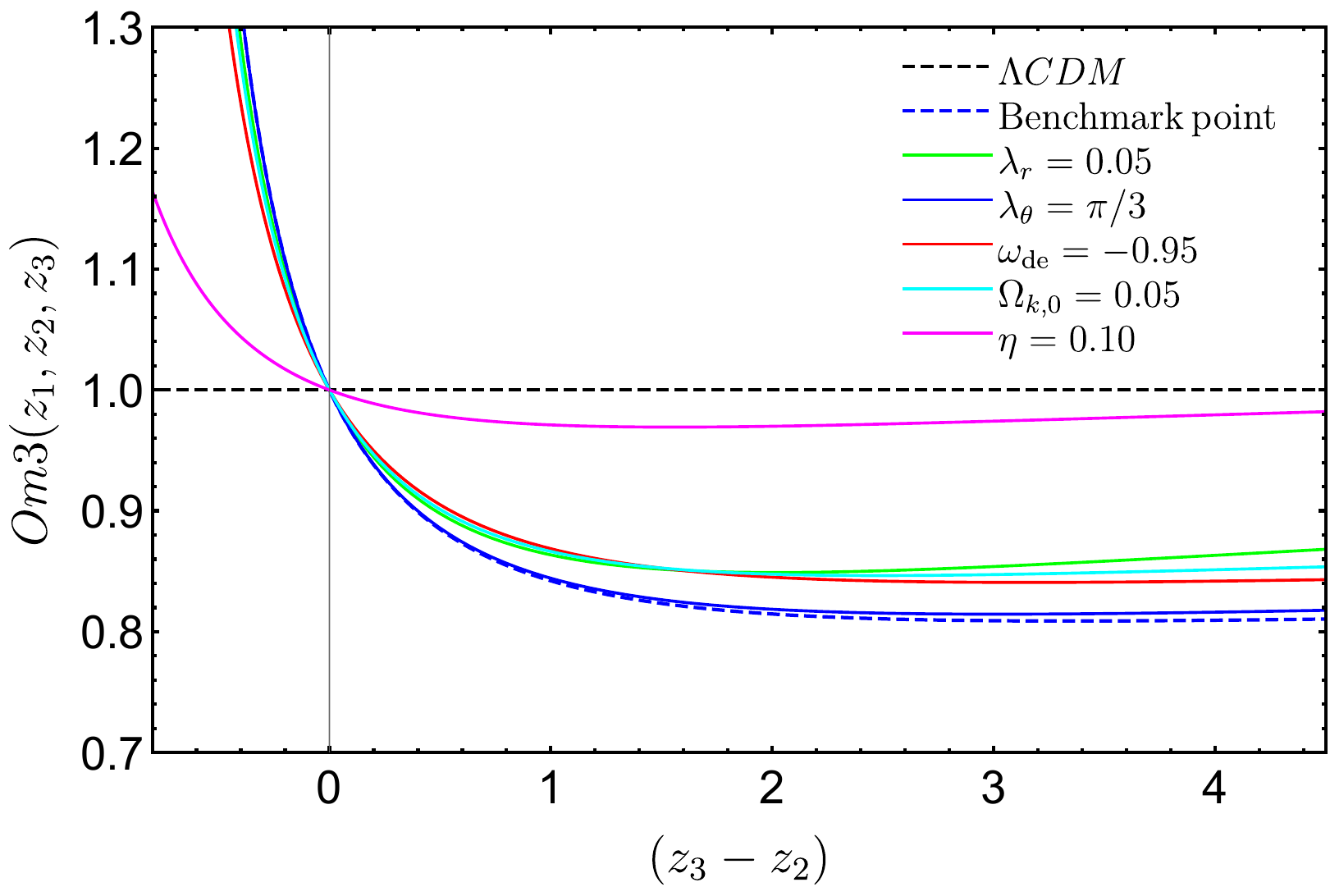}\\
			(a)&(b)\\
		\end{tabular}
		\caption{\label{fig:Omp}(a) Evolution of $Om$ parameters with redshift $z$ for different model parameters. (b) Evolution of $Om3(z_1,z_2,z_3)$ parameter with redshift difference $(z_3-z_2)$ for different model parameters with $z_1=0.20$ and $z_2=0.57$. See text for detail.}
	\end{figure*}
	In order to distinguish the model adopted in this work from the model inspired by $\Lambda$CDM it may be worthwhile to express the $Om3$ parameter for the model. The $Om$ parameter has been extensively studied in the earlier works \cite{10.1093/mnras/stt1941}. The $Om$ parameter is generally defined in terms of hubble parameter in a way that can directly be measured in cosmological observations.
	The $Om$ parameter can be defined as a two point diagnostic given by \cite{shafieloo,sahni}
	\begin{equation}
		Om(z_2,z_1)=\dfrac{h^2(z_2)-h^2(z_1)}{(1+z_2)^3-(1+z_1)^3},
	\end{equation} 
	with $z_1=0$, the $Om$ parameter reduces to 
	\begin{equation}
		Om(z,0)=Om(z)=\dfrac{h^2(z)-1}{(1+z_2)^3-1}.
	\end{equation}
	In general, when $z_1<z_2$, $Om(z_1,z_2)>0$
	
	with $x=z+1$ and $h(x)=H(x)/H_0$, it can be shown that, for $\Lambda$CDM $Om(x)=\Omega_{\rm m,0}$. Also $Om(x)>\Omega_{\rm m,0}$ for quintessence and $Om(x)<\Omega_{\rm m,0}$ for phantom. In terms of $z_1$ and $z_2$ for $\Lambda$CDM, $Om(z_1,z_2)=\Omega_{\rm m,0}$. When $z_1<z_2$, the value of  $Om(z_1,z_2)$ is greater that zero (case of quintessence), whereas $Om(z_1,z_2)<0$ represents the case of phantom. Thus the $Om$ parameter can be useful in distinguishing the DE models. Now the three point diagnostic $Om3$ is defined as \cite{sahni}	
	\begin{equation}
		Om3(z_1,z_2,z_3)=\dfrac{Om(z_2,z_1)}{Om(z_3,z_1)}=\dfrac{h^2(z_2)-h^2(z_1)}{h^2(z_3)-h^2(z_2)} \dfrac{(1+z_3)^3-(1+z_1)^3}{(1+z_2)^3-(1+z_1)^3}.
	\end{equation}
	For $\Lambda$CDM model $Om3=1$ for any redshift $z$. In \autoref{fig:Omp}(a) we show the variation of $Om(z)$ with redshift $z$. In this case $z_1=0$ has been chosen. In this plot $\Lambda$CDM is shown in black dashed line. As expected this is a straight line parallel to $z$ axis at $Om(z)=\Omega_{m,0}\sim 0.31$. The other cases represent the DE model considered here but with different parameters. It appears that all the cases are close to each other except for the case where the viscosity $\eta=0.10$, which differs from the from the rest to a considerable extent. Note that, for other cases,the viscosity is fixed at a value $\eta=0.30$. It appears that as the fluid gets more viscous, the $Om$ parameter tends to deviate more than the $\Lambda$CDM case. It can also be noted from \autoref{fig:Omp}(a) that with the increasing $\lambda_r$ (DM - DE interaction strength) the model deviates more from the $\Lambda$CDM model.
	
	In \autoref{fig:Omp}(b), we plot $Om3(z_1,z_2,z_3)$ with $(z_3-z_2)$ while $z_1$, $z_2$ are kept fixed. The value of $z_1$, $z_2$ are fixed at $z_1=0.20$ \cite{blake}, $z_2=0.57$ \cite{sanchez} respectively and $z_3$ is allowed to vary. From \autoref{fig:Omp}(b) it is seen that the all variations pass through the point where $z_3=z_2$ and at that point $Om3=1$. The deviation of $Om$ from $1$ to $<1$ indicates the phantom nature of dark energy. From this figure (\autoref{fig:Omp}(b)) in can also be observed that with the progress of Universe's evolution, as redshift decreases, the dark energy tends to represent the quintessence case.
	
\begin{figure*}
		\centering
		\begin{tabular}{cc}
			\includegraphics[width=0.49\textwidth]{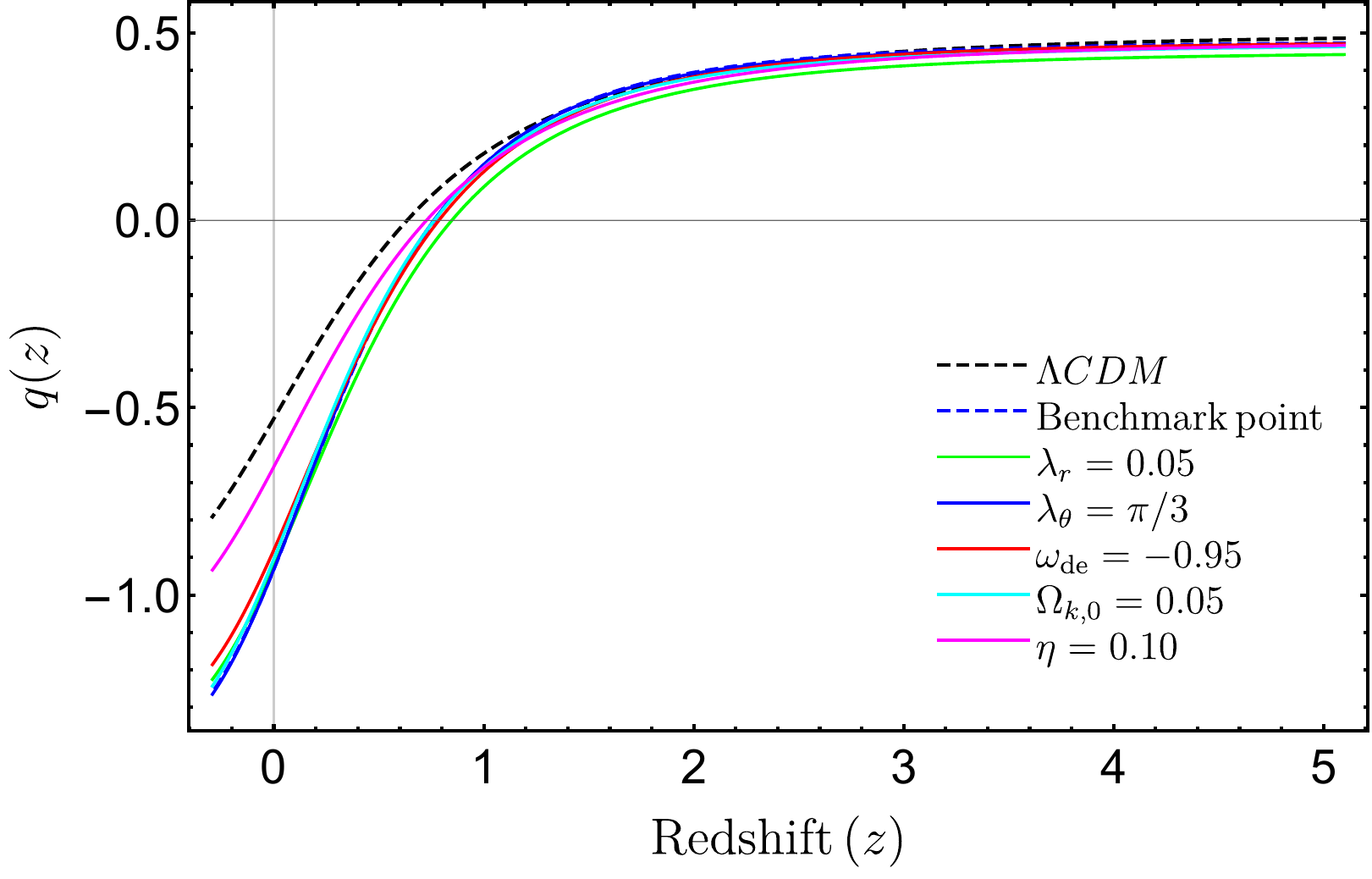}&
			\includegraphics[width=0.49\textwidth]{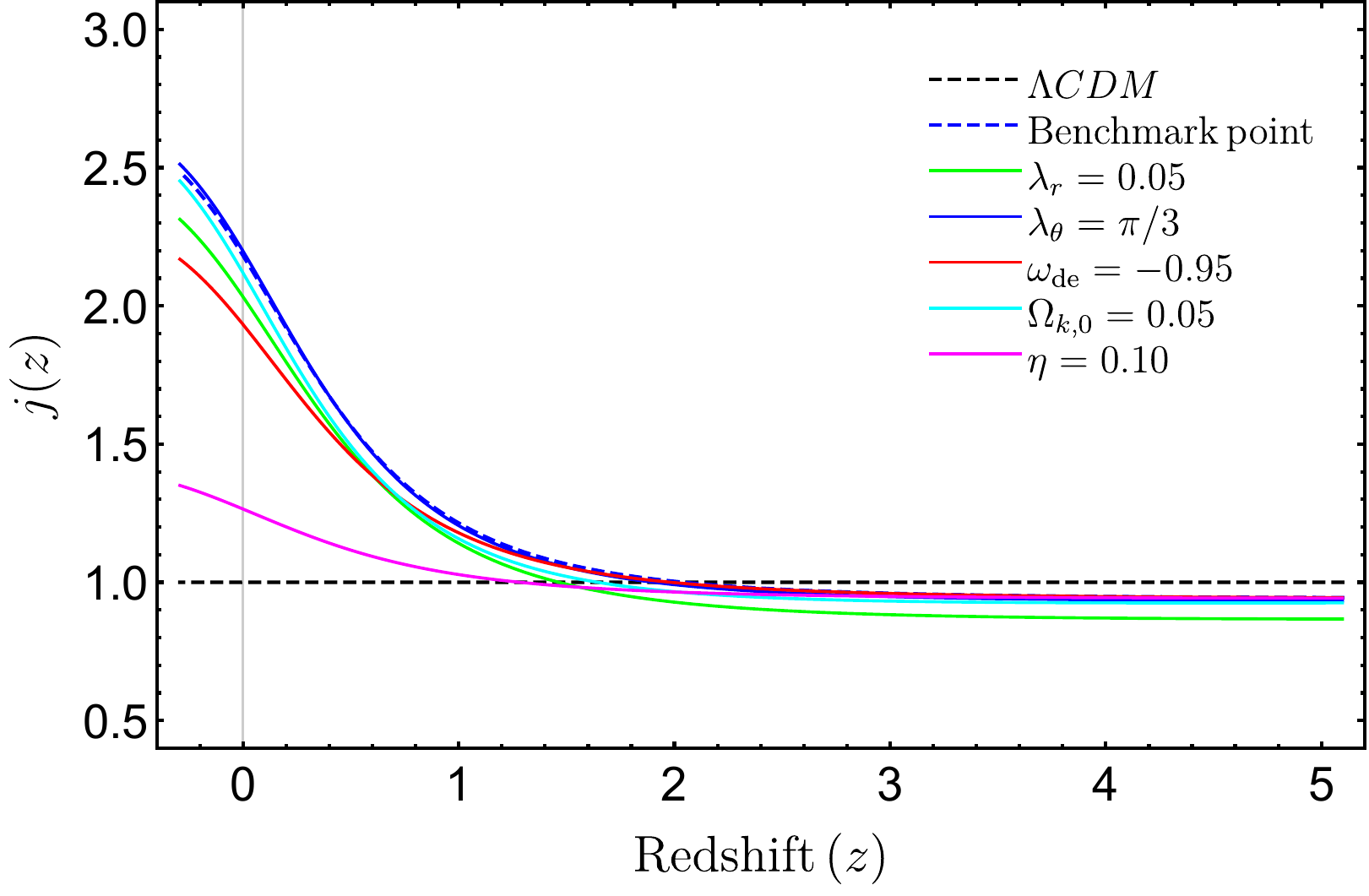}\\
			(a)&(b)\\
		\end{tabular}
		\begin{tabular}{c}
			\includegraphics[width=0.49\textwidth]{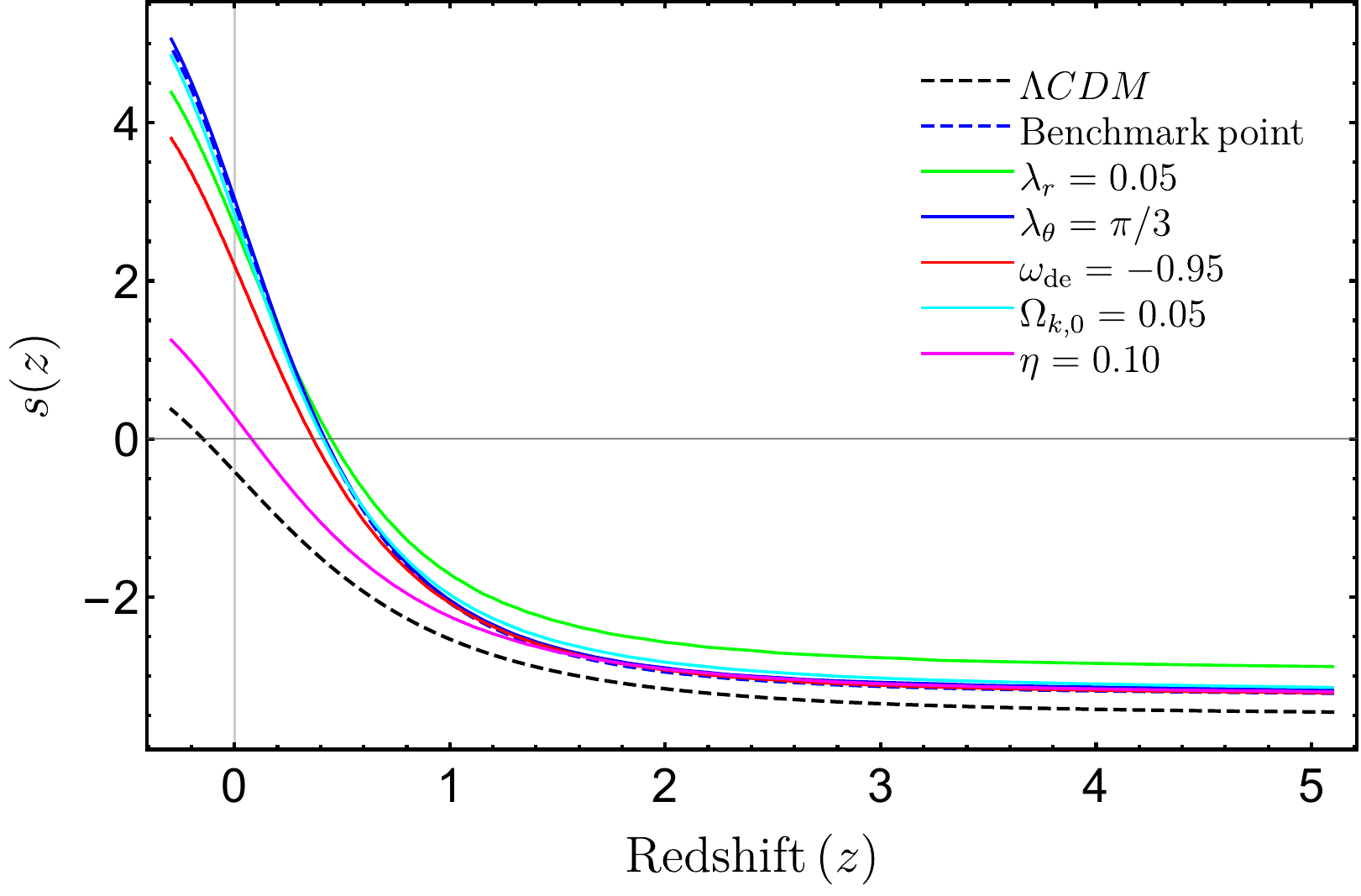}\\
			(c)\\
		\end{tabular}
		\caption{\label{fig:hps} Variation of (a) $q$, (b) $j$ and (c) $s$ parameters with redshift $z$ for different sets of model parameters}
	\end{figure*}

	In the analysis depicted in \autoref{fig:hps}(a), the newly proposed model consistently exhibits a lower value of $q(z)$ compared to the $\Lambda$CDM model, as obtained from the plots. Consequently, the graphs illustrating the jerk parameter $j(z)$ (\autoref{fig:hps}(b)) and snap parameter $s(z)$ (\autoref{fig:hps}(c)) depict higher values relative to the standard cosmic model ($\Lambda$CDM). Notably, the line generated with benchmark values closely aligns with the solid blue line in the plot of deceleration parameter ($q(z)$), although a small discrepancies observed in higher-order parameters ($j(z)$ and $s(z)$). From this figure (\autoref{fig:hps}) one can investigate the impact of $\lambda_r$ in the dynamics of the Universe, which essentially represents the strength of the DM - DE interaction. It is observed that adjustments in $\lambda_r$ contribute to an increase in $q$ both at present and in future cosmic dynamics, while concurrently leading to lower values of $j(z)$ and $s(z)$ across the redshift range. The equation of state parameter for dark energy $\Omega_{\rm de}$ significantly influences the behavior of cosmological parameters. Variations in $\Omega_{\rm de}$ lead to discernible effects on $q(z)$ as well as $j(z)$ and $s(z)$, underscoring the pivotal role of dark energy in shaping expansion dynamics. The effect of the universal curvature is similar to the same for $\omega_{\rm de}$, but lower in magnitude. The curvature of the Universe is represented in this analysis by the present-day curvature density parameter $\Omega_{k,0}$. The effect of the viscous parameter $\eta$ reveals significant influence on cosmological parameters. The magenta line denotes the case where the value of $\eta$ is modified to $\eta=0.10$ from the benchmark set of parameters. It can be observed that the lower value of $\eta$ provides higher values of deceleration and lower values of jerk and snap parameter ($j(z)$ and $s(z)$ respectively).
	
	\begin{figure*}
		\centering
		\begin{tabular}{cc}
			\includegraphics[trim={50 0 0 0},clip,width=0.49\textwidth]{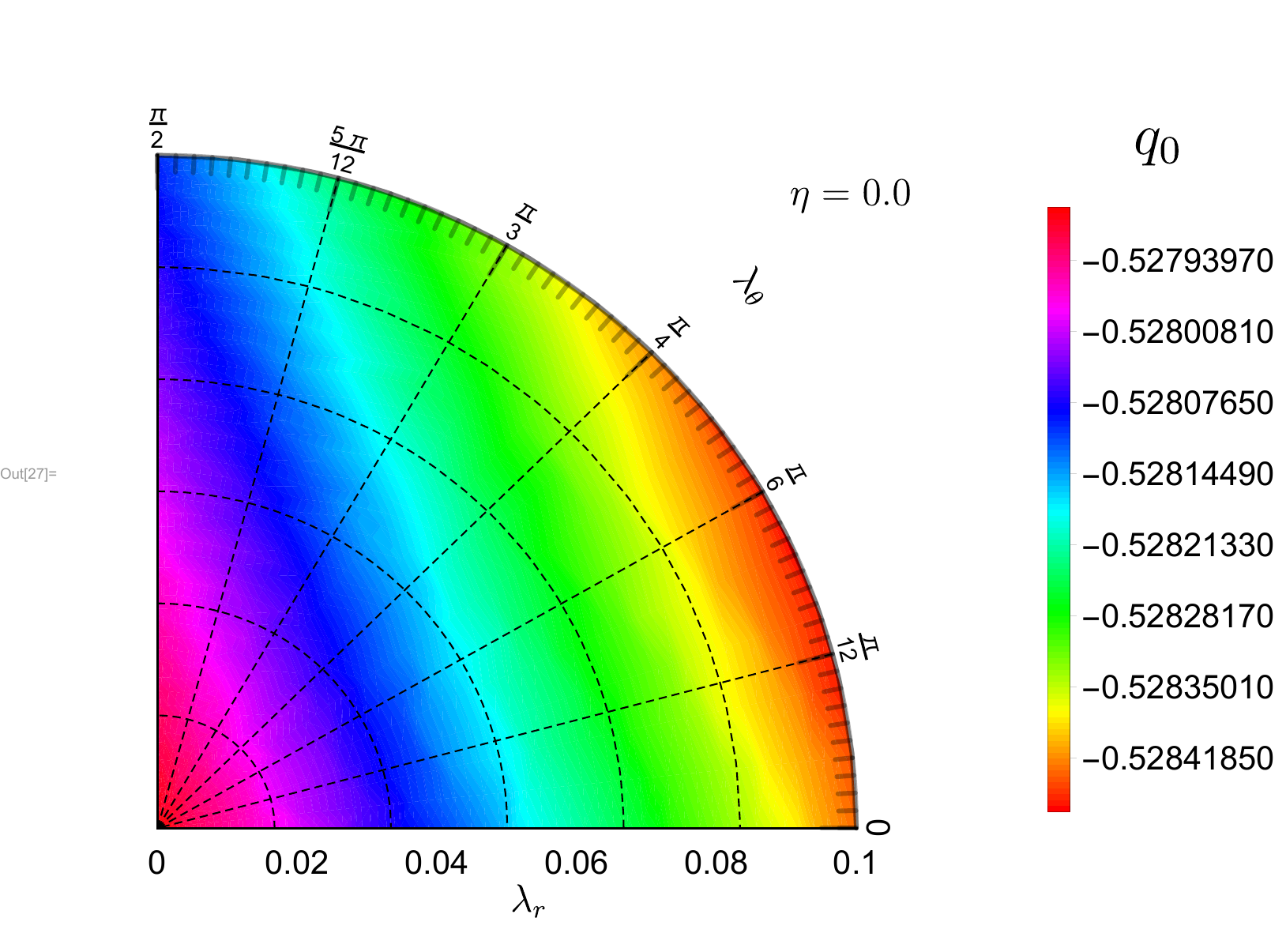}&
			\includegraphics[trim={50 0 0 0},clip,width=0.49\textwidth]{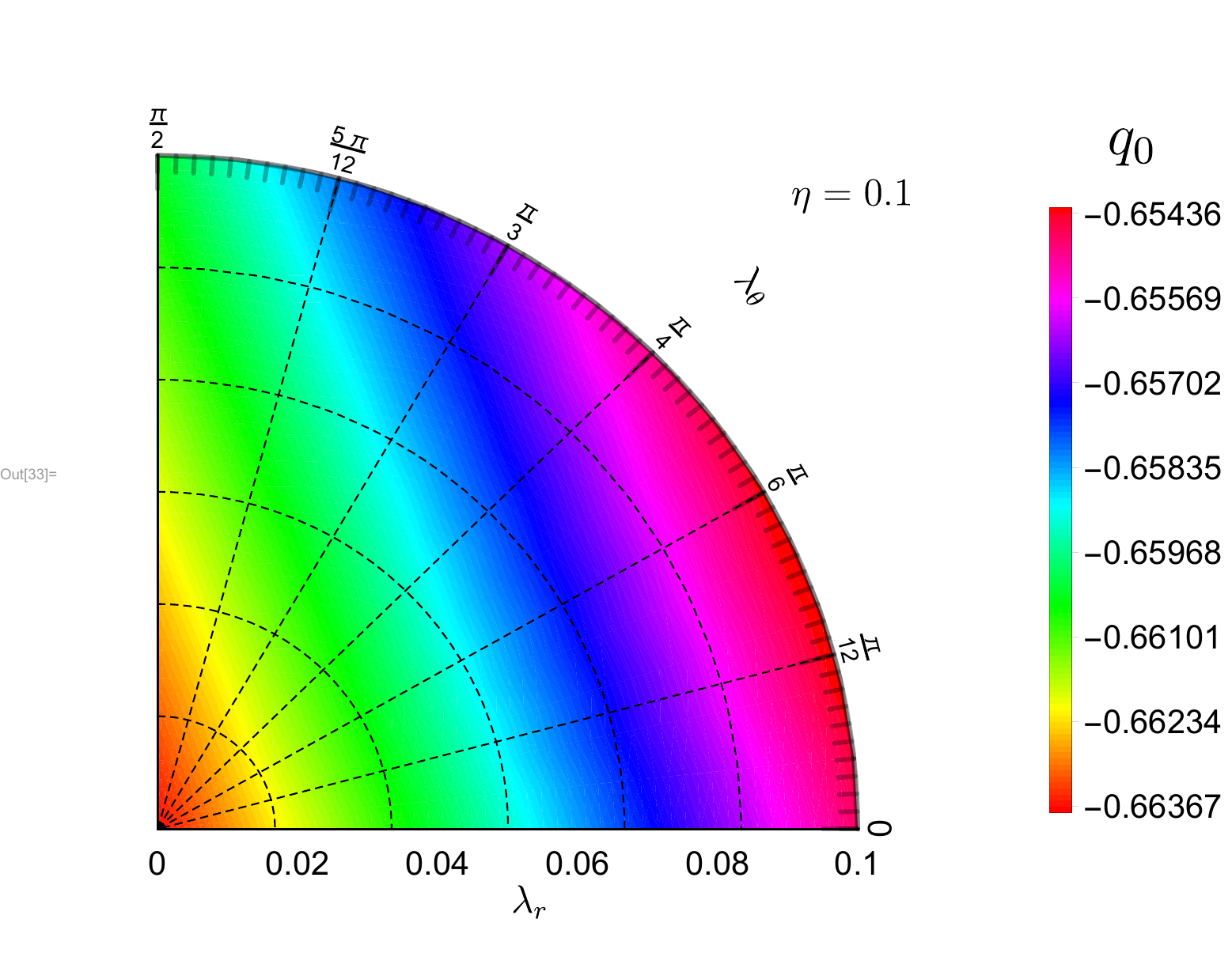}\\
			(a)&(b)\\
			\includegraphics[trim={50 0 0 0},clip,width=0.49\textwidth]{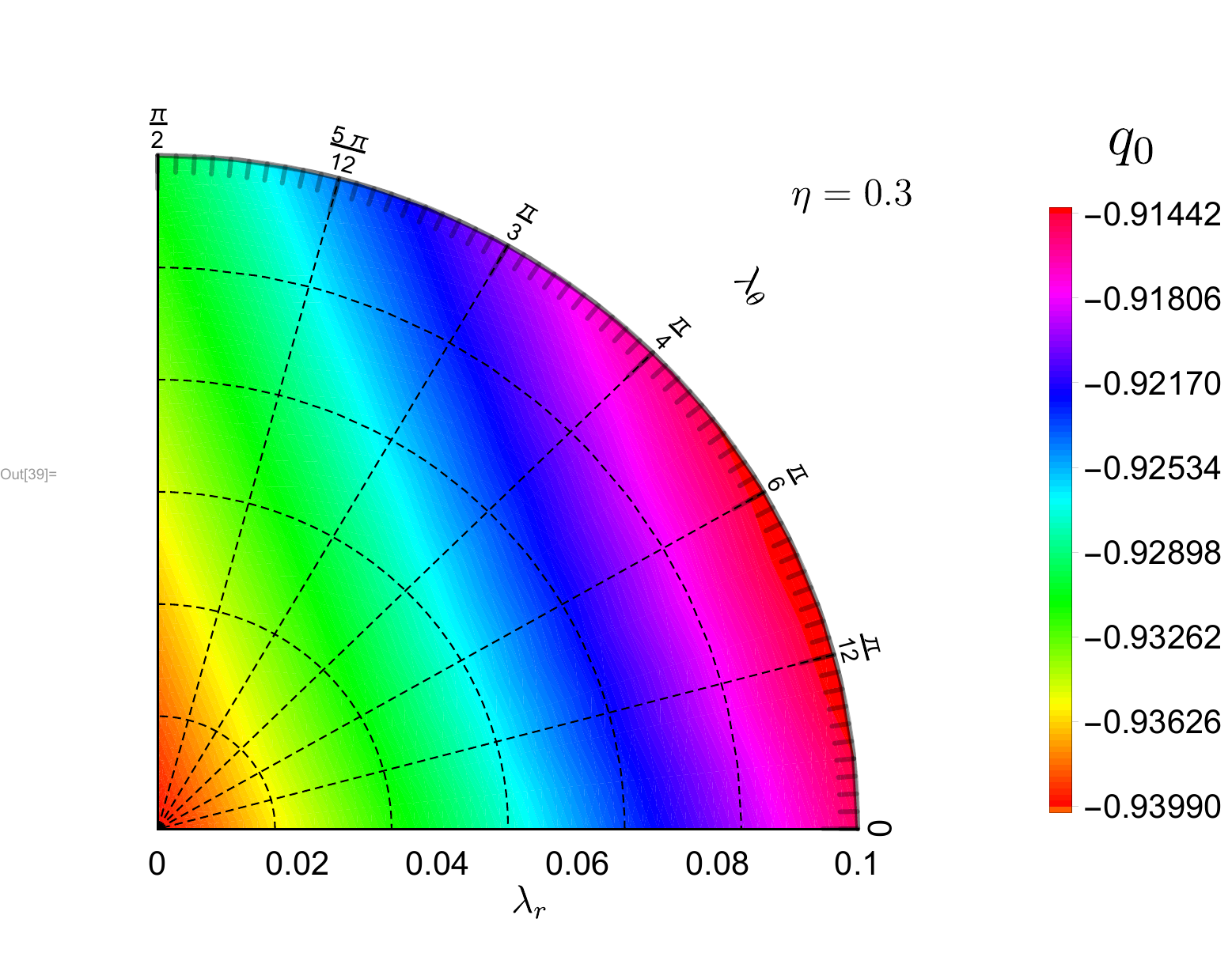}&
			\includegraphics[trim={50 0 0 0},clip,width=0.49\textwidth]{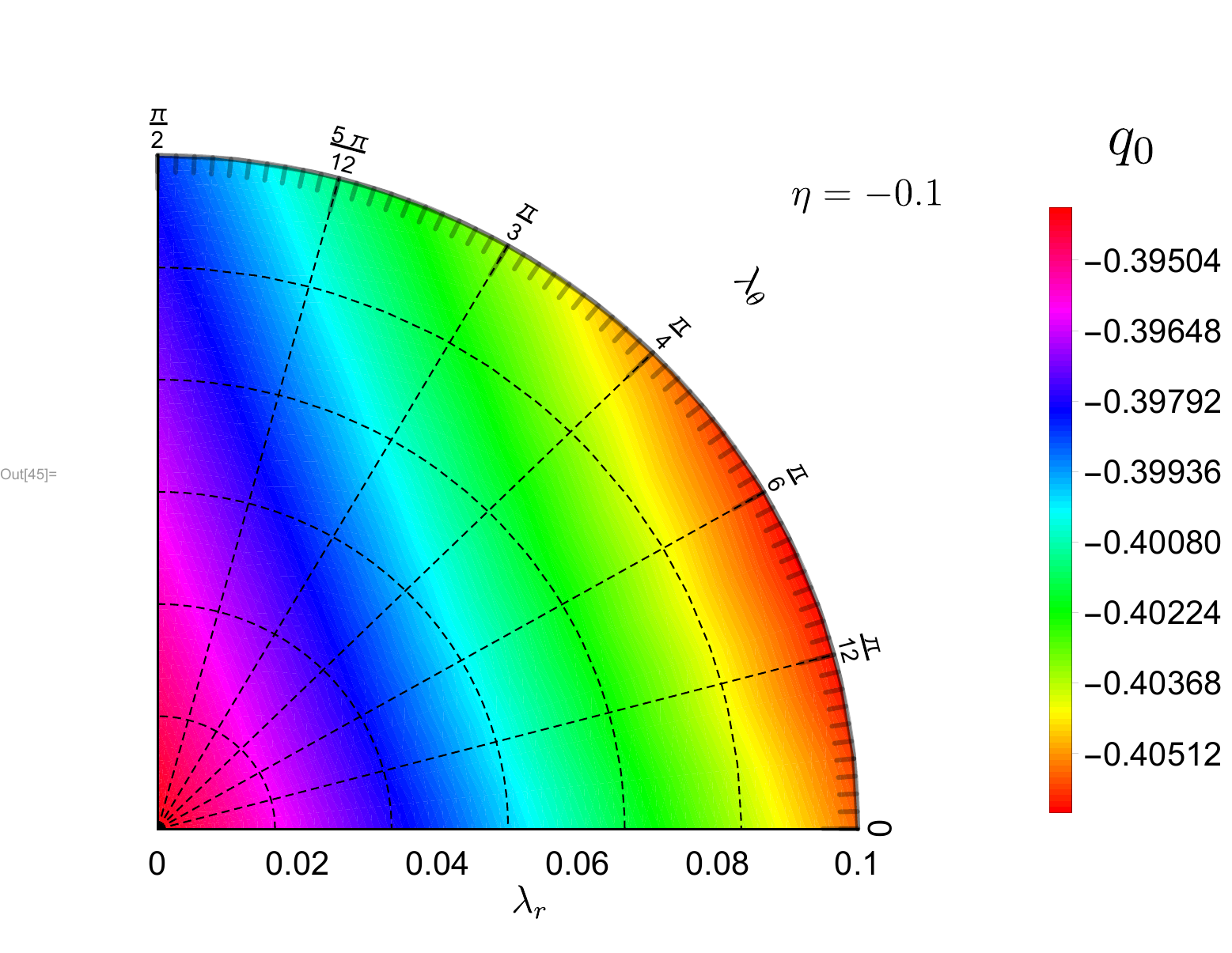}\\
			(c)&(d)\\
		\end{tabular}
		\caption{\label{fig:eta}(a) Variation of deceleration parameter ($q_0$) in $\lambda_r-\lambda_{\theta}$ parameter plane for $\eta=0.0$, (b) for $\eta=0.1$, (c) for $\eta=0.3$, (d) for $\eta=-0.1$ keeping other parameters fixed at $h_0=0.674$, $\Omega_{k,0}=0$ and $\omega_{\rm de}=-1.0$.}
	\end{figure*}
	In the subplots of \autoref{fig:eta}, we visually represent the present deceleration parameter $q_0$ across the $\lambda_r - \lambda_{\theta}$ parameter plane for different values of $\eta$: $\eta=0.0$, $\eta=0.1$, $\eta=0.3$, and $\eta=-0.1$, corresponding to \autoref{fig:eta}(a), \autoref{fig:eta}(b), \autoref{fig:eta}(c), and \autoref{fig:eta}(d), respectively. A noticeable trend emerges wherein, for any given $\eta$, the fluctuation of $q_0$ with respect to $\lambda_r$ is most within the range $\pi/12 \leq \lambda_{\theta} \leq \pi/6$. This fluctuation diminishes for both lower and higher values of $\lambda_{\theta}$ ($\lambda_{\theta} \leq \pi/12$ and $\lambda_{\theta}\geq \pi/6$), reaching a minimum at $\pi/2$.
	
\begin{figure}
		\centering
		\includegraphics[width=0.48\textwidth]{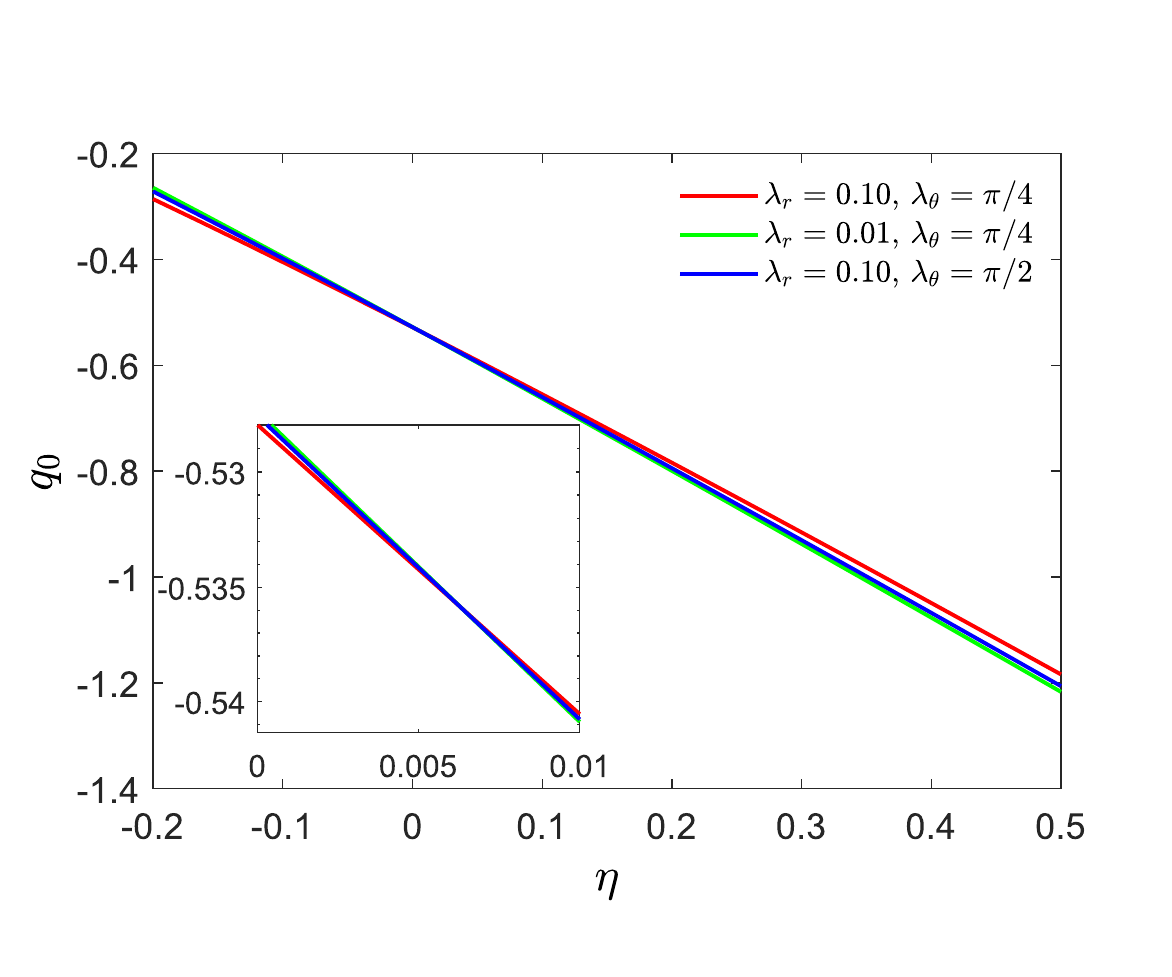}
		\caption{\label{fig:q0} Variation of deceleration parameter ($q_0$) with $\eta$ for various chosen values of $\lambda_r$ and $\lambda_{\theta}$.}
	\end{figure}

	The plots reveal that for higher values of $\eta$ (e.g., $\eta=0.3$ in \autoref{fig:eta}(c)), the fluctuation of $q_0$ is more and $q_0$ increases with increasing $\lambda_r$. However, the resulting range of $q_0$ ($\sim-0.9$) deviates significantly from the same for the case of $\Lambda$CDM model. Conversely, as lower values of $\eta$ are considered, the numerical value of $q_0$ increases, approaching $\sim -0.5$ for $\eta \sim 0.0$. Furthermore, our analysis extends to negative values of $\eta$, where $q_0$ increases further for more negative chosen values of $\eta$ ($\lessapprox -0.52$). An additional phenomena can also be observed that, at lower values of $\eta$, the fluctuation of $q_0$ diminishes significantly, completely disappears around $\eta\approx 0.006$ (see \autoref{fig:q0}). Further reduction in $\eta$ prompts $q_0$ to decrease with increasing $\theta_r$ and the rate of this inverse fluctuation amplifying as $\eta$ decreases even further.
	
	The variation of $q_0$ with $\eta$ is illustrated in \autoref{fig:q0} for different chosen sets of $\lambda_r$ and $\lambda_{\theta}$. From \autoref{fig:q0} it can be seen that, the values of $q_0$ decreases with increasing $\eta$. Although for every set of $\lambda_r$ and $\lambda_{\theta}$, the values of $q_0$ are almost same, the values of $q_0$ falls more rapidly with $\eta$ at higher values of $\lambda_r$. This nature also can be noticed in \autoref{fig:eta}. 
	
	\begin{figure}
		\centering
		\includegraphics[width=\textwidth]{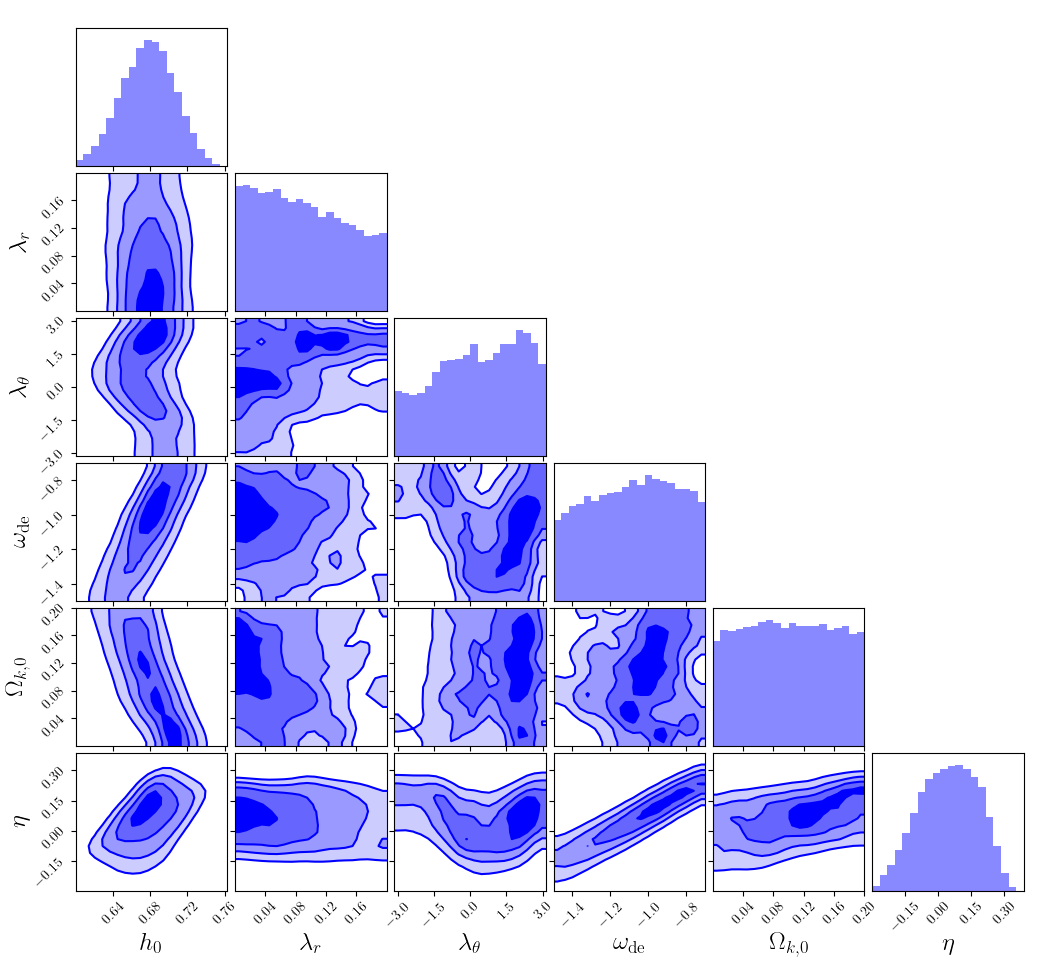}
		\caption{\label{fig:mcmc} Corner plot showing the MCMC result for our model carried out using the observational results of Union 2.1 supernova Ia data. The histograms on the diagonal show the marginalized posterior densities for each parameter.}
	\end{figure}
	The optimum values from the posterior plots are $h_0=0.68$, $\lambda_r=0.09$, $\lambda_{\theta}=0.41$, $\omega_{\rm de}=-1.07$, $\Omega_{k,0}=0.10$ and $\eta=0.05$. However it can be seen that, for the case of $\lambda_r$ the population increases with approaching lower values of $\lambda_r$. On the other hand, $\lambda_{\theta}$ is an angle ranges $-\pi$ to $\pi$, so in this case the case of optimum point is not significant, here the most probable value of $\lambda_{\theta}\approx 7 \pi/10$. 
	
	We conduct a Bayesian analysis to compare our model with the Union 2.1 supernova Ia data. To perform this comparison, we calculate the distance modulus using standard cosmological distance relations incorporating the proposed cosmological model. This calculated distance modulus is then compared with the same as obtained from the Union 2.1 supernova Ia data. In our analysis, the posterior distributions of different parameters are obtained using the Markov Chain Monte Carlo (MCMC) iteration method (\autoref{fig:mcmc}) via the MCMCSTAT package \cite{mcmc1,mcmc2}. We utilize a total of $10^5$ events with an adaptation interval of 500, within the parameter ranges: $h_0 \in [0.6,0.8]$, $\lambda_r \in [0.0,0.2]$, $\lambda_{\theta} \in [-\pi,\pi]$, $\omega_{\rm de} \in [-1.5,-0.7]$, $\Omega_{k,0} \in [0.0,0.2]$ and $\eta \in [-0.3,0.5]$.
	
	The topmost plots in the first, second, third, and fourth columns of \autoref{fig:mcmc} represent the posterior distribution for the parameters $h_0$, $\lambda_r$, $\lambda_{\theta}$, $\omega_{\rm de}$, $\Omega_{k,0}$ and $\eta$ respectively. The remaining plots in \autoref{fig:mcmc} show contour representations of the posterior distribution in various two-parameter spaces. In these contour plots, darker regions denote higher posterior probabilities, and the lines indicate the boundaries of the $1\sigma$, $2\sigma$, and $3\sigma$ regions.
	
	From this analysis, the obtained set of optimum points are $h_0 = 0.68^{+0.03}_{-0.03}$, $\lambda_r = 0.09^{+0.07}_{-0.06}$, $\lambda_{\theta} = 0.41^{+1.86}_{-2.07}$, $\omega_{\rm de} = -1.07^{+0.25}_{-0.27}$, $\Omega_{k,0} = 0.10^{+0.07}_{-0.07}$ and $\eta = 0.05^{+0.13}_{-0.14}$. From the histogram corresponds to $\Omega_{k,0}$ (in \autoref{fig:mcmc}), it can be seen that, the fluctuation of population with $\Omega_{k,0}$ is not very prominent, the pick in the histogram plot for $\Omega_{k,0}$ is also small. So one can conclude that the effect of curvature term is not significant in the dynamics of the Universe where our proposed model is considered. A similar characteristic can be observed in the 1-D histogram plot for $\omega_{\rm de}$, although the pick is prominent in this particular case. On the other hand, the mixing angle $\lambda_{\theta}$ can be varied within $\pi$ to $\pi$, in this case the pick is located at $\approx 7 \pi/10$. The histogram plot corresponds to $\lambda_r$ clearly indicates that the lower value of $\lambda_r$ fits well with the observational data. The posterior distribution for $h_0$ and $\eta$ are resemble to the Gaussian distribution. The optimum values of $h_0$ and $\eta$ are respectively $h_0=0.68$ and $\eta=0.05$. 
	
	\section{Summery and Conclusion} \label{sec:conc}
	
	In this work, using a viscous dark energy framework, we examine how the interaction of dark matter and dark energy affects the evolution of the universe. We present a generalized interacting dark energy (IDE) model that can adjust to different IDE scenarios by varying its parameters, $\lambda_r$ and $\lambda_\theta$, instead of depending on traditional IDE models. Using this generalized model, we analyze the evolution of the Hubble parameter, incorporating nonzero bulk viscosity for dark energy. We begin with discussing three established viscous dark energy scenarios. Following this, we adopt a generalized form of the Hubble parameter for the VDE scenario, as introduced in one of our previous works \cite{21cm_vde}, and eventually modify it to account for interacting dark energy scenarios.
	
    We show the evolution of the Hubble parameter with redshift $z$ under different VDE and IDE scenarios and the fluctuation from the same as calculated for $\Lambda$CDM model in \autoref{fig:first}. This figure demonstrates that higher values of viscosity slow down the expansion, suggesting that viscous effects in DE may act as resistance to acceleration. Interaction between dark matter and dark energy further modifies the expansion rate.
	
	A comparison of the evolution of cosmic density parameters in a VDE and interacting DM - DE model is drawn in \autoref{fig:Omega}(a). In \autoref{fig:Omega}(a), the dark matter density parameter, $\Omega_{\chi}$, the dark energy density parameter, $\Omega_{\rm de}$, the baryon density parameter, $\Omega_{\rm b}$, and the radiation density parameter, $\Omega_{\rm rad}$, are shown with different benchmark parameters. The plot reveals that, while the density parameters for different cosmic components are similar across models at the current epoch. However differences from the $\Lambda$CDM model grow at higher redshifts. It can be observed that an increased value of viscosity parameter $\eta$ helps to deviate the density parameters more from the same as calculated for $\Lambda$CDM model. 
	
	In \autoref{fig:Omega}(b), a parametric representation of the dark energy density parameter $\Omega_{\rm de}$ and total matter density parameter, $\Omega_{\rm m}$ (sum of $\Omega_{\chi}$ and $\Omega_{\rm b}$), shows a straight line trend for $\Lambda$CDM model. In contrast, the inclusion of dark sector interactions results in curved lines and the curvature essentially sensitive to the bulk viscous parameter $\eta$.
	
	\autoref{fig:Omp}(a) shows the variation of $Om(z)$ with redshift $z$. $Om$ parameter for $\Lambda$CDM model is shown by a dashed black line, which remains almost unchanged at $Om(z)=0.31$. However, a significant deviation is observed for the case of higher viscous dark energy. From \autoref{fig:Omp}(a), it is also observed that, an increasing value of $\lambda_r$ results in more pronounced deviations from $\Lambda$CDM.
	
	The three-point diagnostic parameter $Om3(z_1,z_2,z_3)$ is plotted with varying $z_3$ in \autoref{fig:Omp}(b), keeping $z_1=0.20$ and $z_2=0.57$ fixed. For the $\Lambda$CDM model, the parameter $Om3$ remains consistently at unity across all redshifts. From \autoref{fig:Omp}(b), one can see that, all curves pass through $Om3=1$ at $z_3=z_2$, however the value of $Om3$ parameter lie less than unity at $z_3 > z_2$. Here the values of $Om$ parameter less than unity suggests a phantom-like dark energy behavior, while decreasing redshift (progression toward the present Universe) trends toward a quintessence-like nature of dark energy ($Om3>1$), which can be seen at $z_3<z-2$.
	
	In \autoref{fig:hps}(a), our proposed model demonstrates consistently lower values of the deceleration parameter $q(z)$ relative to the $\Lambda$CDM model, while the jerk parameter $j(z)$ and snap parameter $s(z)$ exhibit higher values. The benchmark parameters (dashed blue line) align closely with the deceleration parameters calculated with a fixed value of $\eta$, although slight deviations appear for the case of $j(z)$ and $s(z)$. From (\autoref{fig:hps}) one can analyze the impact of the interaction strength $\lambda_r$ on the Universe's evolution. Increases in $\lambda_r$ result in higher $q$ values, both presently and in future scenarios, while lowering $j(z)$ and $s(z)$ across the redshift range. In contrast, a completely opposite phenomena is observed at $z \gtrapprox0.4$. Furthermore, the dark energy equation of state parameter $\omega_{\rm de}$ significantly influences the evolution of $q(z)$, $j(z)$, and $s(z)$, emphasizing dark energy's central role in cosmic expansion dynamics. The curvature term's influence  ($\Omega_{k,0}$) is minimal to that of $\omega_{\rm de}$ but with reduced magnitude. Modifying the viscous parameter $\eta$ reveals a substantial effect on all parameters. The magenta line in \autoref{fig:hps}(a) shows that lowering $\eta$ to $0.10$ enhances deceleration, while $j(z)$ and $s(z)$ are reduced for the same.
	
	The subplots in \autoref{fig:eta} graphically depicts the value of present deceleration parameter $q_0$ in the $\lambda_r - \lambda_{\theta}$ parameter plane for different vaues of $\eta$: $\eta = 0.0$, $0.1$, $0.3$, and $-0.1$. There is a tendency that the variation of $q_0$ with regard to $\lambda_r$ is most noticeable within $\pi/12 \leq \lambda_{\theta} \leq \pi/6$. Both at lower and higher values of $\lambda_{\theta}$, th fluctuation decreases and it reached its minimum value at $\lambda_{\theta} \approx \pi/2$. 
	
	\autoref{fig:eta}(c) shows that with higher $\eta$ values, as in $\eta = 0.3$, $q_0$ increases with increasing $\lambda_r$. On the other hand, as a lower value of $\eta$ is considered, the variation of $q_0$ with $\lambda_r$ decreases and from $\eta \gtrapprox 0.006$ (see \autoref{fig:q0}) it starts showing reverse variation. It is also observed that, at lower values of $\eta$ correspond to higher $q_0$ values, approaching $-0.5$ for $\eta \approx 0.0$. 
	
	A Bayesian analysis is performed to compare the proposed model with Union 2.1 supernova Ia dataset. Using Markov Chain Monte Carlo (MCMC) sampling with 100,000 events, the posterior distributions for parameters namely $h_0$, $\lambda_r$, $\lambda_{\theta}$, $\omega_{\rm de}$, $\Omega_{k,0}$, and $\eta$ were obtained within their specified ranges. The MCMC results highlighted the optimal parameter values: $h_0 = 0.68$, $\lambda_r = 0.09$, $\lambda_{\theta} = 0.41$, $\omega_{\rm de} = -1.07$, $\Omega_{k,0} = 0.10$, and $\eta = 0.05$. Posterior distributions in \autoref{fig:mcmc} showed that variations in the curvature term $\Omega_{k,0}$ have minimal effect on cosmic dynamics, whereas parameters like $\lambda_{\theta}$ and $\omega_{\rm de}$ show prominent peaks, indicating significant influence. Therefore, in the presence of viscosity in DE fluid, the evolution of Universe has been significantly influenced by DM - DE interaction.
	
	\section*{Acknowledgements}
	
	Two of the authors (A.H. and R.B.) wish to acknowledge the support received from St. Xavier’s College, Kolkata.  One of the authors (R.B.) also thanks the Women Scientist
	Scheme-A fellowship (SR/WOS-A/PM-49/2018), Department of Science and Technology (DST), Govt. of India, for providing financial support.
	
	\bibliography{PUB}
\end{document}